\documentclass[reprint,aps,prl,superscriptaddress]{revtex4-2}
\usepackage{amsmath,amssymb,amsfonts} % mathematics
\usepackage{graphicx} % Include figure files
\usepackage{dcolumn} % Align table columns on decimal point
\usepackage{bm} % bold math
\usepackage[bookmarks=true,colorlinks=true,citecolor=myblue,linkcolor=myblue,filecolor=myblue,urlcolor=myblue]{hyperref} % hyperlink with color
\usepackage{braket} % braket notation in quantum mechanics
\usepackage{tikz} % diagrams in equation
\usepackage{comment} % comment out
\usepackage[normalem]{ulem} % erase
\usepackage{booktabs}
\usepackage{orcidlink}

\def \nr {{n_\mathrm{R}}}
\def \nl {{n_\mathrm{L}}}

\definecolor{mygreen}{RGB}{0,220,0}
\definecolor{myblue}{RGB}{0, 170, 212}

\begin{document}

%%%%%%%%%%%%%%%%%%%%%%%%%%%%%%%%%%%%%%%%%%%%%%%%%%%%%%
%%%%%%%%% title and author information %%%%%%%%%%%%%%%
%%%%%%%%%%%%%%%%%%%%%%%%%%%%%%%%%%%%%%%%%%%%%%%%%%%%%%
% title
\title{Solvable toy model of negative energetic elasticity}

% author information
\author{Atsushi Iwaki\,\orcidlink{0000-0002-5345-8045}}
\email{atsushi-iwaki@phys.s.u-tokyo.ac.jp}
\affiliation{Department of Physics, University of Tokyo, 7-3-1 Hongo, Bunkyo-ku, Tokyo 113-0033, Japan}

\author{Soshun Ozaki\,\orcidlink{0000-0003-4497-7131}}
\affiliation{Department of Basic Science, University of Tokyo, 3-8-1 Komaba, Meguro-ku, Tokyo 153-0041, Japan}

\date{\today}

%%%%%%%%%%%%%%%%%%%%%%%%%%%%%%%%%%%%%%%%%
%%%%%%%%%%%%%% abstract %%%%%%%%%%%%%%%%%
%%%%%%%%%%%%%%%%%%%%%%%%%%%%%%%%%%%%%%%%%
\begin{abstract}
Recent experiments have established negative energetic elasticity, 
the negative contribution of energy to the elastic modulus, 
as a universal property of polymer gels. 
To reveal the microscopic origin of this phenomenon, 
Shirai and Sakumichi investigated a polymer model on a cubic lattice
with the energy effect from the solvent
in finite-size calculations
[\href{https://journals.aps.org/prl/abstract/10.1103/PhysRevLett.130.148101}
{Phys. Rev. Lett. \textbf{130}, 148101 (2023)}]. 
Motivated by this work, 
we provide a simple platform to study
% the elasticity of polymer chains 
negative energetic elasticity
by considering a one-dimensional random walk with the energy effect. 
This model can be mapped onto the classical Ising chain, 
leading to an exact form of the free energy in the thermodynamic or continuous limit. 
Our analytical results are qualitatively consistent with Shirai and Sakumichi's results. 
Our model serves as a fundamental benchmark for studying 
the elasticity of polymer chains. 
% negative energetic elasticity.
\end{abstract}

\maketitle

%%%%%%%%%%%%%%%%%%%%%%%%%%%%%%%%%%%%%%%%%
%%%%%%%%%%%%%%%%%%%%%%%%%%%%%%%%%%%%%%%%%
\emph{\textbf{Introduction.---}}
In condensed matter physics, 
solvable toy models play a crucial role in understanding the essence of phenomena. 
For instance, entropic elasticity, 
which explains the elastic modulus of rubbers, 
appears in a one-dimensional (1D) random walk. 
This serves as a concrete example to comprehend 
how the number of possible states is converted into elasticity 
through statistical mechanics 
\cite{kittel1969, kubo1988}. 

Rubberlike materials consist of polymer chains 
that form entangled and crosslinked networks. 
% This structure provides them with their characteristic elasticity, 
% making them useful in our daily lives. 
The elastic modulus of these materials is determined 
not only by the entropic contribution but also by the energetic one. 
Experiments with natural and synthetic rubbers have demonstrated 
that the energetic contribution is negligibly small 
\cite{meyer1935, anthony1942, mark1976}. 
Consequently, 
the energy effect has been ignored 
in the well-known theories of rubber elasticity 
\cite{flory1953, james1953, flory1969, flory1977}. 
%The small energetic elasticity has also been theoretically explained
%by the rotational isomeric state model \cite{flory1969}. 
However, 
recent experiments have revealed a significant negative energetic contribution in polymer gels, 
which are composed of polymer networks containing a large amount of solvent 
\cite{yoshikawa2021, sakumichi2021, fujiyabu2021, aoyama2023, tang2023-1, tang2023-2}. 
Understanding the microscopic origins of negative energetic elasticity
remains an important problem. 

To address this issue, 
several approaches have been proposed. 
%%%
Shirai and Sakumichi investigated 
a 3D self-avoiding walk, 
a random walk where the overlap with itself is prohibited, 
by conducting an exact enumeration \cite{shirai2023}. 
The self-avoiding walk is acknowledged as an effective lattice model 
for a single polymer chain in a dilute solution 
since it reproduces the excluded volume phenomena of polymers 
\cite{madras1996}. 
The energy effect from solvents was initially discussed in Ref.~\cite{orr1947} 
and later evolved into the so-called interacting self-avoiding walk \cite{vanderzande1998}. 
%This model with self-attractive interactions has been examined 
%in the context of the collapse transition
%\cite{rapaport1974, marenduzzo2003, kumar2007, lee2012, hsieh2016}. 
Shirai and Sakumichi treated the self-repulsive interactions of this model 
and derived negative energetic elasticity. 
%%%
In another research, 
Bleha and Cifra employed the Monte Carlo method 
to study a continuum wormlike chain \cite{bleha2022}, 
which represents the polymer chain as a continuous curve 
and the integral of the curvature gives the bending energy \cite{kratky1949}. 
This model describes semiflexible polymers
characterized by high energy costs for bending 
such as a double-stranded DNA \cite{broedersz2014}. 
They found a negative energetic contribution to force at moderate extension, 
which decreases with higher extension. 
%%%
Additionally, 
Hagita \textit{et al.} conducted
all-atom molecular dynamics simulations 
on polyethylene glycol hydrogels \cite{hagita2023}, 
which was used in the first observation of negative energetic elasticity
\cite{yoshikawa2021}. 
They performed these simulations under both constant-volume and constant-pressure conditions, 
and confirmed that the former shows stronger negativity in energetic elasticity.
%%%
As a more recent work, 
Duarte and Rizzi proposed a simple model, 
a 1D random walk with independent energy at each step \cite{duarte2023}. 
This model is trivially solvable because it is \textit{non-interacting}. 

\begin{figure}
    \centering
    \includegraphics[width=0.8\hsize]{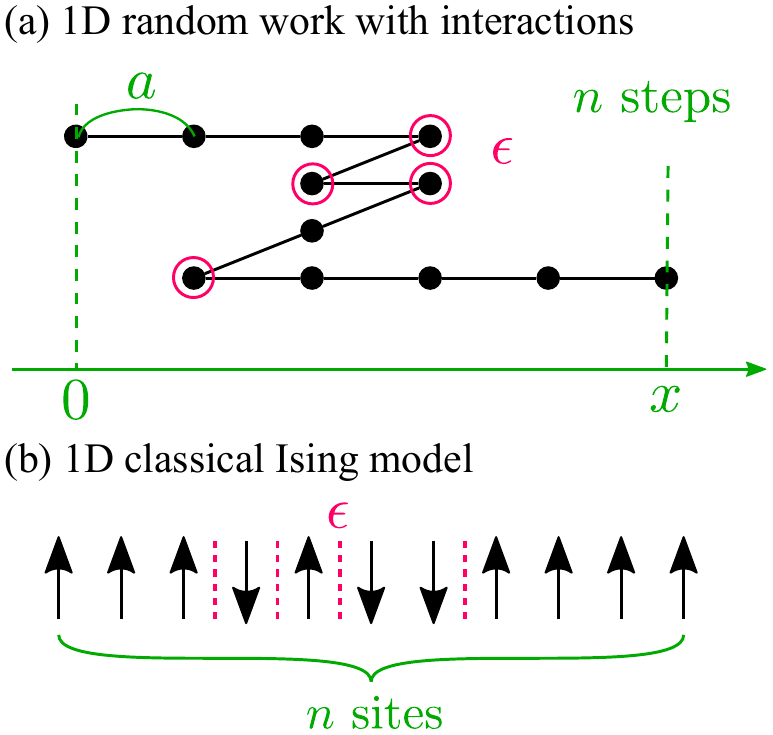}
    \caption{
    Schematic illustrations for 
    (a) a single realization of a 1D random walk 
    with energy cost $\epsilon$ at each bending and 
    (b) the corresponding configuration of the 1D classical Ising model. 
    }
    \label{fig:model}
\end{figure}

There is a need for an \textit{interacting} model 
that can be analytically solved in the thermodynamic limit 
because previous approaches rely on 
finite-size computations, 
numerical simulations, 
or the absence of the interaction. 
In this Letter,
we propose a simple toy model to study negative energetic elasticity
by introducing an energy cost to each bend in a 1D random walk 
as shown in Fig.~\ref{fig:model}(a). 
This model can be mapped onto the 1D classical Ising model
depicted in Fig.~\ref{fig:model}(b), 
resulting in the exact solution 
in the thermodynamic or continuous limit. 
The energy cost at each bending behaves as a self-repulsive interaction 
like Shirai and Sakumichi's work on the self-avoiding walk \cite{shirai2023}.
Our new model successfully explains several important features found 
in previous studies of negative energetic elasticity despite its simplicity.
Our results are a natural extension of entropic elasticity in a 1D random walk, 
thus we believe that this model will be fundamental 
to understand the elasticity of polymer gels. 

%%%%%%%%%%%%%%%%%%%%%%%%%%%%%%%%%%%%%%%%%
%%%%%%%%%%%%%%%%%%%%%%%%%%%%%%%%%%%%%%%%%
\emph{\textbf{Model.---}}
We start with
a 1D random walk 
characterized by the number of steps $n$, 
the width per step $a$, 
and the final position $x$ as shown in Fig.~\ref{fig:model}(a). 
To incorporate the energy effect, 
we introduce an energy term $\epsilon$ 
for each bending. 
% a step changing the direction. 
For instance, the realization depicted in Fig.~\ref{fig:model}(a)
incurs an energy cost of $4\epsilon$. 
By associating right (left) steps with up (down) spins, 
this model can be mapped onto the 1D classical Ising model 
as shown in Fig.~\ref{fig:model}(b)
The Hamiltonian of this model is given by
\begin{align}
    H_0 (\bm{\sigma}) 
    &= \sum_{i=1}^{n-1} \epsilon 
    \frac{1- \sigma_i \sigma_{i+1}}{2} \notag \\
    &= - \frac{\epsilon}{2} \sum_{i=1}^{n-1} \sigma_i \sigma_{i+1}
    + \epsilon \frac{n-1}{2}, 
\end{align}
where $\sigma_i = \pm 1$ is the classical spin at site $i$, 
corresponding to the right/left direction of the $i$-th step.  
With an external force $F_\mathrm{ex}$, 
the whole Hamiltonian becomes 
\begin{align}
    H(\bm{\sigma} ; F_\mathrm{ex}) 
    &=  H_0(\bm{\sigma}) - F_\mathrm{ex} a \sum_{i=1}^n \sigma_i \notag \\
    &= - \frac{\epsilon}{2} \sum_{i=1}^{n-1} \sigma_i \sigma_{i+1}
    - F_\mathrm{ex} a \sum_{i=1}^n \sigma_i
    + \epsilon \frac{n-1}{2}, 
    \label{eq:ham_fex}
\end{align}
because the final position $x$ is expressed by the spins 
as $x = a \sum_{i=1}^n \sigma_i$. 
These Hamiltonians are the 1D classical Ising model 
with and without a magnetic field up to an energy constant. 
Our model is a 1D version of a discrete wormlike chain \cite{rosa2003} 
which is equivalent to the classical XY model. 

In the following, 
we exactly calculate the thermodynamic functions 
under two independent conditions; 
the external force $F_\mathrm{ex}$ is fixed in the first 
and the final position $x$ is fixed in the second. 
The first case is described by the Ising model in a magnetic field, 
whereas the second case is described by the Ising model with a fixed magnetization. 
The force-fixed and position-fixed conditions correspond to 
the constant-pressure and constant-volume condition 
of all-atom molecular dynamics in Ref.~\cite{hagita2023}, 
respectively. 
Under both conditions, 
we can obtain the relationship between $x$ and $F_\mathrm{ex}$ 
by differentiating the thermodynamic functions. 
This relationship allows us 
to calculate the stiffness $k= \partial F_\mathrm{ex} / \partial x$, 
which represents the elastic modulus of a single chain, 
including its energetic and entropic contributions. 

%%%%%%%%%%%%%%%%%%%%%%%%%%%%%%%%%%%%%%%%%
%%%%%%%%%%%%%%%%%%%%%%%%%%%%%%%%%%%%%%%%%
\emph{\textbf{Force-fixed condition.---}}
The force-fixed condition is represented 
by the Hamiltonian in Eq.~\eqref{eq:ham_fex}. 
%We need to consider all possible realizations of a random walk with an external force $F_\mathrm{ex}$, 
%which means all configurations of the Hamiltonian described by Eq.~\eqref{eq:ham_fex}.
Consequently, the partition function
at the temperature $k_B T=1/\beta$ is given by
\begin{equation}
    Z_g(\beta, F_\mathrm{ex}, n) 
    = \sum_{\bm{\sigma}} e^{-\beta H(\bm{\sigma} ; F_\mathrm{ex})}. 
\end{equation}
The free energy associated with this partition function can be calculated 
using the traditional transfer-matrix method \cite{baxter1982}. 
The partition function is represented by the transfer matrix as 
\begin{equation}
    Z_g(\beta, F_\mathrm{ex}, n) 
    = \bm{v}^\top X^{n-1} \bm{v}, 
\end{equation}
where
\begin{equation}
    X = 
    \begin{bmatrix}
        e^{\beta F_\mathrm{ex} a} & e^{-\beta \epsilon} \\
        e^{-\beta \epsilon} & e^{-\beta F_\mathrm{ex} a}
    \end{bmatrix}, \quad 
    \bm{v} = 
    \begin{bmatrix}
        e^{\beta F_\mathrm{ex} a /2} \\ e^{- \beta F_\mathrm{ex} a /2} 
    \end{bmatrix}. 
\end{equation}
From the largest eigenvalue of $X$, 
the Gibbs free energy, the thermodynamic function under this condition, 
can be calculated as 
\begin{align}
    &\beta g_\mathrm{th} (\beta, F_\mathrm{ex})
    = - \lim_{n\to\infty} \frac{1}{n} \log Z_g(\beta, F_\mathrm{ex}, n) \notag \\
    &= - \log \left[ \cosh(\beta F_\mathrm{ex} a) 
    + \sqrt{\sinh^2 (\beta F_\mathrm{ex} a) + e^{-2\beta\epsilon}} \right]. 
    \label{eq:gibbs}
\end{align}
The relationships between $F_\mathrm{ex}$ and $x$ is determined 
through $x/n = -\partial g_\mathrm{th} / \partial F_\mathrm{ex}$ as 
\begin{equation}
    \sinh(\beta F_\mathrm{ex} a) 
    = e^{-\beta\epsilon} \frac{y}{\sqrt{1-y^2}},  
    \label{eq:fex_x}
\end{equation}
where $y=x/na$ is the rescaled dimensionless position. 
The stiffness $k = \partial F_\mathrm{ex} / \partial x$ 
can be calculated as a function of $\beta$ and $y$: 
\begin{equation}
    \Hat{\beta} \Hat{k} (\Hat{\beta}, y) 
    = \frac{e^{-\Hat{\beta}}}{1-y^2} 
    \sqrt{\frac{1}{1 + (e^{-2\Hat{\beta}} - 1) y^2}}. 
    \label{eq:stiffness}
\end{equation}
Here, we have introduced rescaled dimensionless quantities 
$\Hat{k} = na^2k/\epsilon$ and $\Hat{\beta} = \beta\epsilon$. 

In order to decompose the stiffness into its energetic and entropic contributions 
as $k = k_U + k_S$ \cite{yoshikawa2021}, 
we consider the Helmholtz free energy $f_\mathrm{th}(\beta, x)$, 
which is the thermodynamic function under the position-fixed condition
calculated in the next part explicitly. 
Since we can obtain $F_\mathrm{ex}$ 
by the differentiation of $f_\mathrm{th}$ 
as $F_\mathrm{ex} /n = \partial f_\mathrm{th} / \partial x$, 
the stiffness is given by $k/n = \partial^2 f_\mathrm{th} / \partial x^2$. 
Considering that $f_\mathrm{th}$ is decomposed into the energetic and entropic term as 
$f_\mathrm{th} = u - Ts$, 
the energetic and entropic contributions to the stiffness can be defined as 
$k_U/n = \partial^2 u / \partial x^2$ and 
$k_S/n = - T~ \partial^2 s / \partial x^2$, respectively. 
By applying Maxwell's relation, 
both contributions can be computed from $k(T, x)$ as 
$k_S = T~\partial k / \partial T$ 
and $k_U = k - k_S$. 

By substituting the equilibrium value $y=0$ when $F_\mathrm{ex}=0$, 
the stiffness and its energetic and entropic contributions 
are derived from Eq.~\eqref{eq:stiffness} as 
\begin{equation}
    \Hat{\beta} \Hat{k} = e^{-\Hat{\beta}}, \quad 
    \Hat{\beta} \Hat{k}_U = - \Hat{\beta} e^{-\Hat{\beta}}, \quad 
    \Hat{\beta} \Hat{k}_S = (1+\Hat{\beta}) e^{-\Hat{\beta}}, 
    \label{eq:betaks}
\end{equation}
where $\Hat{k}_U = na^2k_U/\epsilon$ and $\Hat{k}_S = na^2k_S/\epsilon$ 
are rescaled dimensionless quantities. 
Therefore, when the interactions are repulsive ($\epsilon >0$), 
the energetic contribution becomes negative ($k_U<0$), 
indicating negative energetic elasticity. 
As will be plotted later, 
the behavior of Eq.~\eqref{eq:betaks} is 
qualitatively consistent with Fig.~3 of Ref.~\cite{shirai2023} 
on the interacting self-avoiding walk. 

In the vicinity of a temperature $T_0$, 
the stiffness is approximated by 
\begin{equation}
    k(T) \simeq \frac{k_S(T_0)}{T_0}T + k_U(T_0). 
\end{equation}
Therefore, 
in experiments and numerical simulations, 
when the stiffness is lineally approximated as $k(T) \simeq a(T - T_U)$, 
a dimensionless quantity
\begin{equation}
    \Hat{T}_U = \frac{k_B T_U}{\epsilon} 
    = - \frac{k_U(T_0)}{k_S(T_0)} \frac{k_B T_0}{\epsilon}
\end{equation}
is used as an indicator of negative energetic elasticity. 
We can analytically obtain $\Hat{T}_U$ in the limit of $T_0 \to \infty$ as 
\begin{equation}
    \Hat{T}_U^\infty = \lim_{T_0 \to \infty} \Hat{T}_U(T_0)
    = 1 - y^2, 
    \label{eq:tuinf}
\end{equation}
indicating that negative energy elasticity diminishes 
as the chain is extended. 
This result is consistent with the previous work 
on the wormlike chain \cite{bleha2022}. 

Although the equilibrium value is $y=0$ when $F_\mathrm{ex}=0$, 
the chain is considered to be extended by thermal fluctuations. 
Here, we investigate the mean square of the position $\langle y^2 \rangle$. 
Using the Gibbs free energy in the thermodynamic limit, 
the mean square is calculated as 
\begin{align}
    \langle y^2 \rangle
    = - \frac{1}{n} \frac{\partial^2 g_\mathrm{th}}{\partial F_\mathrm{ex}^2} (\beta, F_\mathrm{ex}=0)
    = \frac{e^{\beta\epsilon}}{n}. 
\end{align}
Thus, $y=0$ is the stable equilibrium value in the thermodynamic limit $n\to\infty$. 
Since $\langle y^2 \rangle$ is bounded by 1, 
this equality approximately holds if $e^{\beta\epsilon} \ll n$, 
In contrast, if $e^{\beta\epsilon} \gg n$, 
the mean square saturates to $\langle y^2 \rangle \simeq 1$. 
To explore the temperature region 
where $e^{\beta\epsilon}$ and $n$ are comparable, 
we perform a finite-size scaling analysis \cite{antal2004}. 
By fixing $\alpha = n / e^{\beta\epsilon}$ and taking the limit $n\to\infty$, 
we obtain 
\begin{align}
    \langle y^2 \rangle
    &= \frac{e^{-2\alpha} - 1 + 2\alpha}{2\alpha^2}.
    \label{eq:scale_y2}
\end{align}
This form of scaling function explains the behavior mentioned above. 
By taking the same scaling limit, 
we can represent the stiffness as a function of $\alpha$ as 
\begin{align}
    \Hat{\beta} \Tilde{k}
    = \frac{1}{\langle y^2 \rangle}
    = \frac{2\alpha^2}{e^{-2\alpha} -1 + 2\alpha}, 
    \label{eq:scale_k}
\end{align}
where $\Tilde{k} = n^2 a^2 k / \epsilon$ 
is a rescaled quantity for this limit. 
The energetic component of the stiffness 
also conforms to a scaling function as
\begin{align}
    \Tilde{k}_U
    = \frac{\partial}{\partial \Hat{\beta}} \Hat{\beta} \Tilde{k}
    = -4 \alpha^2 
    \frac{(1+\alpha)e^{-2\alpha} - 1 + \alpha}{(e^{-2\alpha} -1 + 2\alpha)^2}, 
    \label{eq:scale_ku}
\end{align}
where $\Tilde{k}_U = n^2 a^2 k_U / \epsilon$ is a rescaled quantity. 
In the regime where $\alpha \ll 1$, 
indicating the chain is fully extended due to thermal fluctuations, 
the stiffness is polynomially suppressed with inverse temperature as 
$\Tilde{k} \simeq 1/ \beta\epsilon$. 
On the other hand, its energetic contribution is exponentially small,  
expressed as $\Tilde{k}_U \simeq - 2n / 3e^{\beta\epsilon}$. 
Consequently, the ratio of the energetic part to the stiffness 
decreases with a chain extension, 
aligning with Eq.~\eqref{eq:tuinf}. 

%%%%%%%%%%%%%%%%%%%%%%%%%%%%%%%%%%%%%%%%%
%%%%%%%%%%%%%%%%%%%%%%%%%%%%%%%%%%%%%%%%%
\emph{\textbf{Position-fixed condition.---}}
% Next, we consider the situation under the position-fixed condition. 
The position-fixed condition corresponds to the Ising model with a fixed magnetization, 
% This situation corresponds to the Ising model with a fixed magnetization, 
where the partition function is given by 
\begin{equation}
    Z_f(\beta, x, n) = \sum_{\bm{\sigma}}
    e^{-\beta H_0(\bm{\sigma})} 
    \delta\left( \frac{x}{a}, \sum_{i=1}^N \sigma_i \right). 
\end{equation}
We categorize the summation based on the number of bends as
\begin{align}
    Z_f(\beta, x, n)= \sum_{m} W(m, x, n) e^{-\beta\epsilon m},
\end{align}
where $W(m, x, n)$ is the number of cases of $m$-bending realizations
with the final position $x$ of a $n$-step random walk. 
Since $W(m, x, n)$ does not depend on the sign of $x$, 
we have $Z_f(\beta,x, n)=Z_f(\beta,-x, n)$. 
Denoting the number of right and left steps as 
\begin{equation}
    \nr = \frac{n}{2} + \frac{x}{2a}, \quad
    \nl = \frac{n}{2} - \frac{x}{2a}, 
\end{equation}
and assuming $x\geq0$, i.e., $\nr\geq\nl$,
$W(m,x,n)$ is evaluated by division into four cases:
the random walks starting with the right/left step and
ending with the right/left step.
Therefore, we obtain the partition function as 
\begin{align}
    Z_f(\beta, x, n) =
    &2\sum_{m=1}^\nl
    \binom{\nr-1}{m-1} \binom{\nl-1}{m-1}
    e^{-(2m-1)\beta \epsilon} \nonumber \\
    &+
    \sum_{m=1}^\nl 
    \binom{\nr-1}{m} \binom{\nl-1}{m-1} 
    e^{-2m\beta \epsilon}\nonumber \\
    &+
    \sum_{m=1}^{\nl-1} 
    \binom{\nr-1}{m-1} \binom{\nl-1}{m}
    e^{-2m\beta \epsilon}. 
\end{align}
The logarithm of this summation can traditionally be evaluated using the maximum-term method
in the thermodynamic limit \cite{hill1960}. 
Instead, we will evaluate the summation by inserting the Kronecker delta 
and obtain an exact integral form of the partition function. 
This approach enables us not only to calculate the free energy in the thermodynamic limit 
but also to account for finite-size correction terms in the numerical part 
and perform the finite-size scaling similar to the previous section. 
First, 
we insert a Kronecker delta in an integral form,
\begin{align}
    \delta_{lm}=\frac{1}{2\pi i} \oint_C \frac{dz}{z} z^{l-m},
\end{align}
where $C$ is a counterclockwise contour around the origin. 
By carrying out the summation, 
we obtain the partition function in an integral form as 
\begin{multline}
    Z_f(\beta, x, n) 
    =\frac{e^{-\beta\epsilon}}{\pi i} 
    \oint_C \frac{dz}{z} 
    (1+ e^{-\beta\epsilon} z)^{\nr-1} \\
     \times (1+ e^{-\beta\epsilon} z^{-1})^{\nl-1} 
    \left( 1+\frac{z+z^{-1}}{2} \right). 
\end{multline}
This representation also holds when $x < 0$ 
because the change of the variable $z\to w=1/z$ 
confirms $Z_f(\beta,x, n)=Z_f(\beta,-x, n)$. 
Applying the saddlepoint approximation, 
we obtain the Helmholtz free energy, 
the thermodynamic function under this condition, as 
\begin{widetext}
\begin{multline}
    \beta f_\mathrm{th}(\beta, y) 
    = - \lim_{n\to\infty} \frac{1}{n} \log Z_f(\beta, nay, n)
    = -\frac{1+y}{2} \log \left[1+ (1-e^{-2\beta\epsilon})y + e^{-\beta\epsilon}\sqrt{1-(1-e^{-2\beta\epsilon})y^2}\right] \\
    -\frac{1-y}{2} \log \left[1- (1-e^{-2\beta\epsilon})y + e^{-\beta\epsilon}\sqrt{1-(1-e^{-2\beta\epsilon})y^2}\right]
    +\frac{1+y}{2} \log(1+y) + \frac{1-y}{2}\log(1-y). 
    \label{eq:helmholtz}
\end{multline}
\end{widetext}
This expression and the relation $F_\mathrm{ex} a = \partial f_\mathrm{th} / \partial y$
also leads to Eq.~\eqref{eq:fex_x}.
This is because
the Helmholtz free energy $f_\mathrm{th}(\beta, y)$ contains the same information 
as the Gibbs free energy $g_\mathrm{th}(\beta, F_\mathrm{ex})$. 
These two thermodynamic functions are connected to each other 
through the Legendre transformation as
\begin{gather}
    g_\mathrm{th}(\beta, F_\mathrm{ex})
    = \min_y \left[ f_\mathrm{th}(\beta, y)- F_\mathrm{ex}ay \right], 
    \label{eq:fth_to_gth} \\
    f_\mathrm{th}(\beta, y) 
    = \max_{F_\mathrm{ex}} 
    \left[ g_\mathrm{th}(\beta, F_\mathrm{ex}) + F_\mathrm{ex} ay \right].
    \label{eq:gth_to_fth}
\end{gather}

%%%%%%%%%%%%%%%%%%%%%%%%%%%%%%%%%%%%%%%%%
%%%%%%%%%%%%%%%%%%%%%%%%%%%%%%%%%%%%%%%%%
\emph{\textbf{Numerical demonstrations.---}}
Here, we examine our analytical results 
and compare them to finite-size numerical results with $n=20$
under the force-fixed and position-fixed conditions. 
In addition, 
we demonstrate the finite-size scaling
with $F_\mathrm{ex}=0$ for various system sizes. 

\begin{figure}
    \centering
    \includegraphics[width=\hsize]{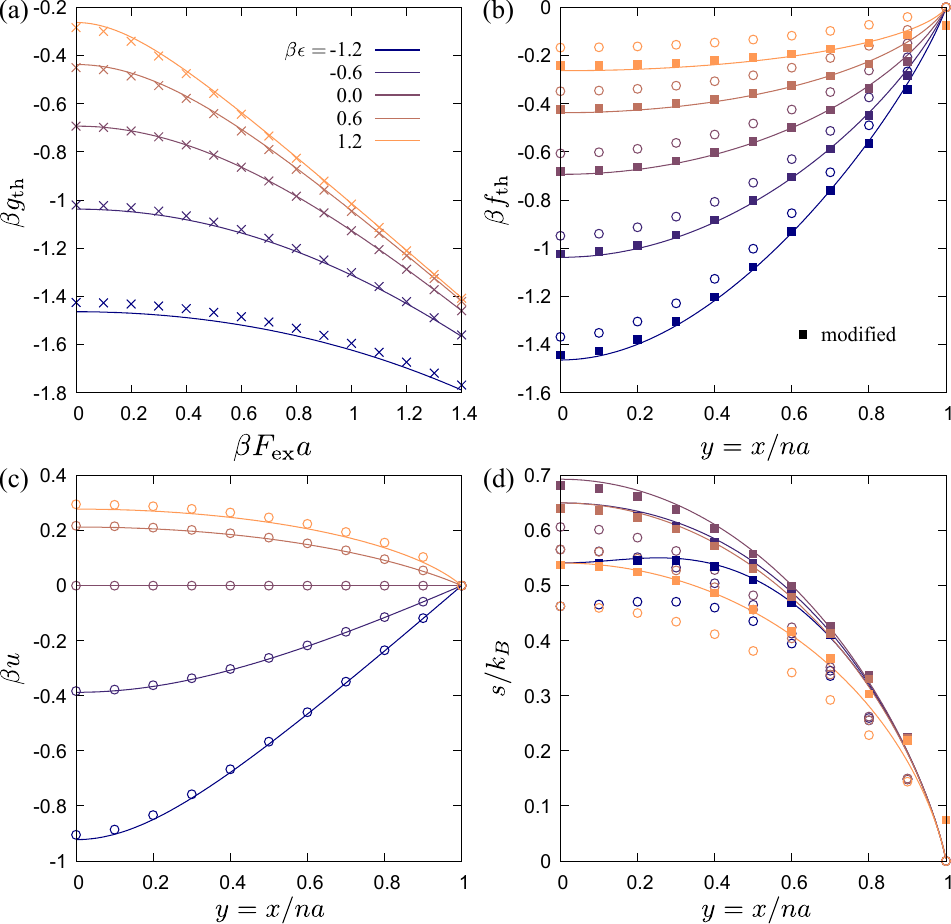}
    \caption{
    (a) Gibbs free energy $\beta g_\mathrm{th}$ 
    under the force-fixed condition in Eq.~\eqref{eq:gibbs} and 
    (b) Helmholtz free energy $\beta f_\mathrm{th}$ 
    under the position-fixed condition in Eq.~\eqref{eq:helmholtz}. 
    (c) Energy $\beta u$ and (d) entropy $s/k_B$ as functions of $y=x/na$ 
    obtained from the Helmholtz free energy $\beta f_\mathrm{th}$. 
    These quantities are computed for $\beta\epsilon = 0, \pm 0.6, \pm 1.2$. 
    Data points represent numerical results with $n=20$. 
    Square points in panels (b) and (d) express modified results 
    to remove the dominant term of the finite-size effect
    in Eq.~\eqref{eq:modification}.
    }
    \label{fig:2}
\end{figure}

Figures~\ref{fig:2}(a) and (b) show the thermodynamic functions
under both conditions in Eqs.~\eqref{eq:gibbs} and \eqref{eq:helmholtz}. 
Under the force-fixed condition, 
the equilibrium value of $F_\mathrm{ex}$ is determined 
by maximizing $g_\mathrm{th} + F_\mathrm{ex}ay$ following Eq.~\eqref{eq:gth_to_fth}. 
Thus, 
the Gibbs free energy $\beta g_\mathrm{th}$ is convex upward as seen in Fig.~\ref{fig:2}(a). 
On the other hand, 
the equilibrium value of $y$ is obtained
by minimizing $f_\mathrm{th}- F_\mathrm{ex}ay$ following Eq.~\eqref{eq:fth_to_gth} 
under the position-fixed condition, 
resulting in a convex downward behavior of the Helmholtz free energy $\beta f_\mathrm{th}$ 
as illustrated in Fig.~\ref{fig:2}(b). 
The data points in Fig.~\ref{fig:2}(a) are consistent with the analytical results. 
However, in Fig~\ref{fig:2}(b), 
the raw numerical data represented by circle points differs from the exact solutions 
due to the finite-size effect. 
Detailed calculation of the saddle point approximation in Eq.~\eqref{eq:helmholtz} 
allows us to evaluate the dominant term of this finite-size effect as 
\begin{equation}
    - \frac{1}{n} \log Z_f(\beta, nay, n) \simeq 
    \beta f_\mathrm{th} (\beta, y) + \frac{1}{2n} \log n. 
    \label{eq:modification}
\end{equation}
This dominant term $\log n /2n$, which is independent of $\beta$ and $y$, 
does not affect energy, external force, and stiffness. 
We modify the numerical results by subtracting the dominant term $\log n/2n$ 
from the finite-size calculations of the free energy. 
The modified numerical results shown in Fig.~\ref{fig:2}(b) with square dots
are in improved agreement with the exact solutions. 

We decompose the Helmholtz free energy $\beta f_\mathrm{th}$ 
into the energy $\beta u$ and entropy $s/k_B$, 
which are shown in Figs.~\ref{fig:2}(c) and (d), respectively. 
Because the energetic and entropic contributions to the stiffness 
are determined by the second derivative of these quantities, 
the convexity observed in Figs.~\ref{fig:2}(c) and (d) characterizes each contribution to elasticity. 
When the interaction is repulsive as $\epsilon > 0$, 
$\beta u$ displays upward convexity, 
indicating negative energetic elasticity. 
In the self-attractive case with $\epsilon < 0$, 
the entropy becomes convex downward near $y=0$ at low temperature, 
which means negative entropic elasticity. 
After removing the finite-size effect according to Eq.~\eqref{eq:modification}, 
the numerical results are in good agreement with the exact solutions. 

\begin{figure}
    \centering
    \includegraphics[width=0.9\hsize]{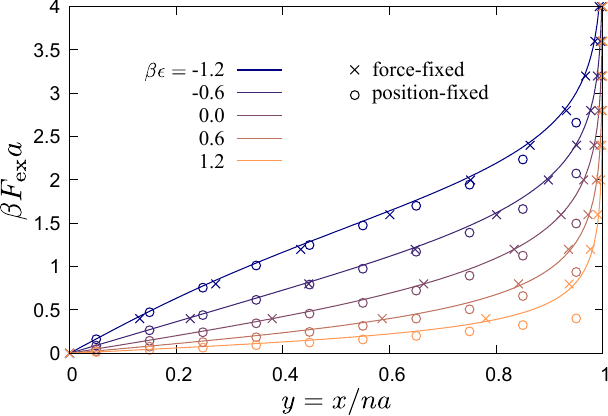}
    \caption{
    Relationship between the external force $\beta F_\mathrm{ex} a$ 
    and the final position $y=x/na$ in Eq.~\eqref{eq:fex_x} 
    for various $\beta\epsilon$. 
    Cross ($\times$) and circle ($\circ$) points represent 
    numerical results with $n=20$ under the force-fixed and position-fixed conditions, 
    respectively. 
    }
    \label{fig:fex_x}
\end{figure}

The relation between the external force $\beta F_\mathrm{ex} a$ 
and final position $y$ in Eq.~\eqref{eq:fex_x} is plotted in Fig.~\ref{fig:fex_x}. 
The numerical data are generated by differentiating the thermodynamic functions 
shown in Figs.~\ref{fig:2}(a) and (b). 
For the position-fixed condition, 
$\beta F_\mathrm{ex} a$ is approximated using a finite difference form as 
\begin{equation}
    \beta F_\mathrm{ex} a \left( y + \frac{\Delta y}{2} \right) 
    \simeq \frac{\beta f_\mathrm{th}(y + \Delta y) - \beta f_\mathrm{th}(y)}{\Delta y}, 
\end{equation}
where $\Delta y = 2 / na$. 
As expected, 
the stronger force is required to extend the chain for the larger $\beta\epsilon$. 
The numerical results under the force-fixed condition show good agreement with the exact solutions across the whole range, 
while discrepancies arise at high extensions under the position-fixed condition. 
This is because the finite-size effect depends on the ensemble used; 
it is exponentially small in the force-fixed condition 
and polynomially small in the position-fixed condition. 

\begin{figure}
    \centering
    \includegraphics[width=0.9\hsize]{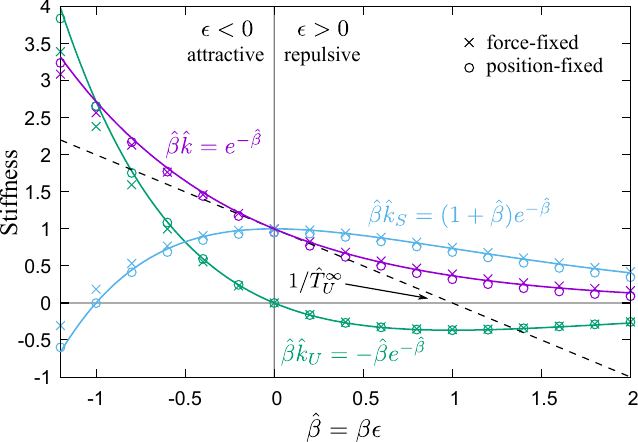}
    \caption{
    Stiffness $\Hat{\beta}\Hat{k}$ and 
    its energetic and entropic contributions, $\Hat{\beta}\Hat{k}_U$ and $\Hat{\beta}\Hat{k}_S$, 
    with the final position $y=0$ in Eq.~\eqref{eq:betaks}. 
    Cross ($\times$) and circle ($\circ$) points represent 
    numerical results with $n=20$ under the force-fixed condition at $F_\mathrm{ex}=0$ 
    and the position-fixed condition at $y=0$, 
    respectively. 
    The tangent line of $\Hat{\beta}\Hat{k}$ at $\Hat{\beta}=0$ depicted by a dashed line 
    intersects the horizontal axis at $1/\Hat{T}_U^\infty$. 
    }
    \label{fig:stiffness}
\end{figure}

Figure~\ref{fig:stiffness} displays the stiffness and its energetic and entropic contributions
at $y=0$ in Eq.~\eqref{eq:betaks}. 
The numerical data are derived from the second derivative of the thermodynamic functions shown in Figs.~\ref{fig:2}(a) and (b)
at $F_\mathrm{ex} = 0$ or $y=0$, 
employing a finite-difference form for the position-fixed case as 
\begin{equation}
    \Hat{\beta}\Hat{k}(y) \simeq 
    \frac{\beta f_\mathrm{th}(y + \Delta y) - 2\beta f_\mathrm{th}(y) + \beta f_\mathrm{th}(y - \Delta y)}
    {(\Delta y)^2}. 
\end{equation}
We observe negative energetic elasticity in the self-repulsive region with $\epsilon > 0$ 
and negative entropic elasticity in the self-attractive ($\epsilon<0$) and low-temperature region, 
which aligns with the convexity seen in Figs.~\ref{fig:2}(c) and (d). 
The dashed line represents the tangent line of $\Hat{\beta}\Hat{k}$ at $\Hat{\beta}=0$. 
This line crosses the horizontal axis at $1/\Hat{T}_U^\infty$, 
which is determined analytically in Eq.~\eqref{eq:tuinf}; 
in this case with $y=0$, $\Hat{T}_U^\infty = 1$. 
The numerical results are almost coincident with the exact solutions for both conditions. 
The qualitative behavior of Fig.~\ref{fig:stiffness} is consistent 
with those reported in Fig.~3 of Ref.~\cite{shirai2023}
on the interacting self-avoiding walk. 

\begin{figure}
    \centering
    \includegraphics[width=0.9\hsize]{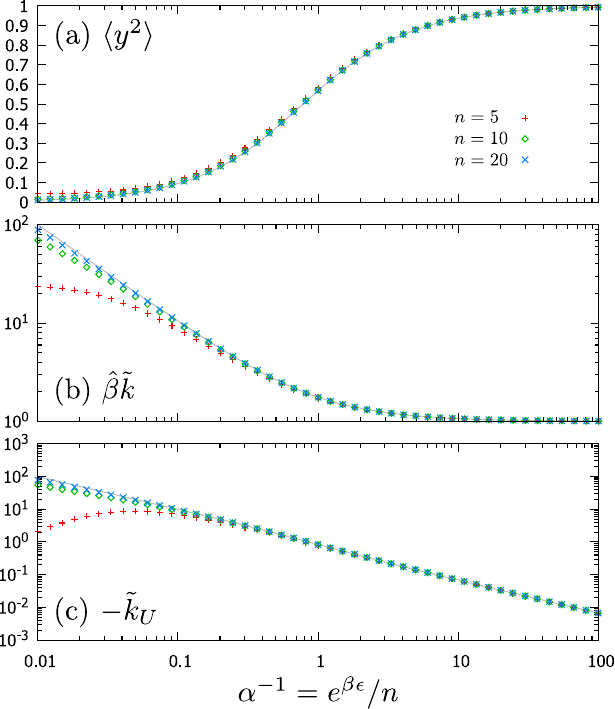}
    \caption{
    (a) The means square of the position $\langle y^2 \rangle$, 
    (b) the stiffness $\Hat{\beta}\Tilde{k}$, and 
    (c) its energetic contribution $-\Tilde{k}_U$ 
    as functions of $\alpha^{-1} = e^{\beta\epsilon} / n$
    for $n = 5, 10, 20$ 
    when $F_\mathrm{ex} = 0$. 
    Solid lines represent the scaling functions 
    in Eqs.~\eqref{eq:scale_y2}, \eqref{eq:scale_k}, and \eqref{eq:scale_ku}. 
    }
    \label{fig:crossover}
\end{figure}

Figure~\ref{fig:crossover} shows the finite-size scaling analysis 
for the mean square position, the stiffness, and its energetic contribution 
under the condition $F_\mathrm{ex} = 0$. 
The numerical result for $n=10$ already closely align 
with the scaling functions described in Eqs.~\eqref{eq:scale_y2}, \eqref{eq:scale_k}, and \eqref{eq:scale_ku} 
across a broad temperature range. 
As illustrated in Fig.~\ref{fig:crossover}(b), 
$\Hat{\beta} \Tilde{k}$ becomes constant 
at high temperature $e^{\beta\epsilon} \gg n$. 
In contrast, 
Fig.~\ref{fig:crossover}(c) demonstrates that 
$\Tilde{k}_U$ continues to decrease over the whole temperature range. 
Consequently, 
the negative energetic elasticity reduces 
as the chain elongates due to thermal fluctuations.

%%%%%%%%%%%%%%%%%%%%%%%%%%%%%%%%%%%%%%%%%
%%%%%%%%%%%%%%%%%%%%%%%%%%%%%%%%%%%%%%%%%
\emph{\textbf{Summary and discussion.---}}
We have proposed a toy model to explore negative energetic elasticity, 
a 1D random walk with an energy cost $\epsilon$ for each bending. 
This model can be mapped onto the 1D classical Ising model 
by associating right (left) steps with up (down) spins. 
Table~\ref{tab:correspondence} shows how these models correspond. 
We have exactly calculated the thermodynamic functions 
under the force-fixed and position-fixed conditions 
in Eqs.~\eqref{eq:gibbs} and \eqref{eq:helmholtz}
using the transfer-matrix formulation and saddlepoint approximation. 
The Legendre transformation connects these thermodynamic functions 
as they are equivalent. 
We analytically obtained the stiffness in Eq.~\eqref{eq:stiffness}, 
which has a negative energetic contribution 
when the interaction is repulsive as $\epsilon > 0$. 

Our model reproduces several properties of previous studies 
on negative energetic elasticity. 
The stiffness and its energetic and entropic contributions 
in the simplest case in Eq.~\eqref{eq:betaks} plotted in Fig.~\ref{fig:stiffness} 
are qualitatively consistent with the results of a 3D self-avoiding walk 
in Fig.~3 of Ref.~\cite{shirai2023}. 
Equation~\eqref{eq:tuinf} shows that 
the negative energetic contribution decreases compared to the entropic one with chain extension, 
in accordance with Ref.~\cite{bleha2022} for a continuum wormlike chain. 
The force-fixed (position-fixed) condition of our model corresponds to 
the constant-pressure (constant-volume) condition 
of all-atom molecular dynamics simulations in Ref.~\cite{hagita2023}. 
Therefore, the finite-size effect in these simulations may resemble our model. 
In this sense, 
our model serves as an essential platform 
for investigating the elasticity of polymer chains. 

In addition to reproducing existing findings, 
we have succeeded in characterizing the behavior of stiffness 
in the regime where thermal fluctuations extend the chain 
as scaling functions in Eqs.~\eqref{eq:scale_k} and \eqref{eq:scale_ku}, 
plotted in Fig.~\ref{fig:crossover}. 
Our work provides a guideline for understanding the stiffness behavior 
resulting from changes in the folding structure of polymers 
which is more effective in higher-dimensional spaces. 

\begin{table}
    \centering
    \caption{
    Correspondence between the 1D random walk with interactions 
    and the 1D classical Ising model. 
    }
    \begin{ruledtabular}
    \begin{tabular}{cc}
        Random walk & Ising model \\
        \midrule
        right/left step & up/down spin \\
        external force $F_\mathrm{ex}$ & magnetic field $h$ \\
        final position $x$ & magnetization $M$ \\
        stiffness $\displaystyle k = \frac{\partial F_\mathrm{ex}}{\partial x}$ & 
        $(\mathrm{susceptibility})^{-1}$ $\displaystyle \chi^{-1} = \left(\frac{\partial M}{\partial h}\right)^{-1}$ \\
    \end{tabular}
    \end{ruledtabular}
    \label{tab:correspondence}
\end{table}

We can extend our model to $d$-dimensional space with $d>1$. 
In this space, steps in each direction are mapped onto one of $2d$ possible states arranged in a line and 
the energy cost for bending becomes interactions between different states adjacent to each other. 
This framework leads to the $2d$-state Potts model in 1D, 
which can also be analytically solved under the fixed-force condition 
using the transfer-matrix method. 

%%%%%%%%%%%%%%%%%%%%%%%%%%%%%%%%%%%%%%%%%
%%%%%%%%%%%%%%%%%%%%%%%%%%%%%%%%%%%%%%%%%
\emph{\textbf{Acknowledgments.---}}
We thank Yuta Sakai for comments on the Ising model.  
We are also grateful to Naoyuki Sakumichi and Hidehiro Saito for fruitful discussions. 
A. I. and S. O. were supported by a Grant-in-Aid for JSPS Research Fellow (Grants No. 22KJ0661 and No. 22KJ0988).

%%%%%%%%%%%%%%%%%%%%%%%%%%%%%%%%%%%%%%%%%%%%
%%%%%%%%%%%%%% bibliography %%%%%%%%%%%%%%%%
%%%%%%%%%%%%%%%%%%%%%%%%%%%%%%%%%%%%%%%%%%%%
\nocite{apsrev42Control}
\bibliographystyle{apsrev4-2}
\bibliography{references}

\end{document}